\documentclass[a4paper,11pt]{article}
\pdfoutput=1 

\usepackage{jheppub} 

\usepackage{tikz}
\usepackage{bm}
\newcommand*{\circled}[1]{\lower.7ex\hbox{\tikz\draw (0pt, 0pt)%
    circle (.5em) node {\makebox[1em][c]{\small #1}};}}

\usepackage{subcaption}

\title{Constraints on the Scale Parameter of Regular Black Hole in Asymptotically Safe Gravity from Extreme Mass Ratio Inspirals}

\author[1]{lai Zhao}
\author[1]{Meirong Tang}
\author[1,*]{Zhaoyi Xu}

\affiliation[1]{College of Physics, Guizhou University, Guiyang 550025, China}

\emailAdd{zyxu@gzu.edu.cn(Corresponding author)}

\abstract{
This paper evaluates the potential for constraining the quantum scale parameter $\xi$ of regular black hole within the asymptotically safe gravity framework using gravitational waves from extreme mass ratio inspirals (EMRIs). Since $\xi$ cannot be precisely determined from first principles, observational constraints become crucial. We employ the Augmented Analytical Kludge (AAK) method to calculate gravitational waveforms in the equatorial plane and systematically analyze the influence of different $\xi$ values on phase evolution. Comparison with the Schwarzschild case demonstrates that the corrective effects of $\xi$ accumulate in the phase over observation time, thereby providing distinguishable observational signatures. Through waveform mismatch analysis, our results indicate that the LISA detector can effectively detect the presence of $\xi$ at the $\sim10^{-4}$ level for systems with a mass of $10^6M_\odot$. Further assessment using the Fisher information matrix (FIM)  confirms a measurement precision of $\Delta\xi\approx3.225\times10^{-4}$, which significantly surpasses existing observational methods, providing quantitative observational evidence for asymptotically safe quantum gravity theory in the strong-field regime.
}


\begin{document}
\maketitle
\flushbottom
\section{\label{sec:level1}Introduction}
Black holes predicted by general relativity have been confirmed through gravitational wave detection \cite{LIGOScientific:2016emj,LIGOScientific:2016aoc,LIGOScientific:2018mvr,LIGOScientific:2020aai} and Event Horizon Telescope observations \cite{EventHorizonTelescope:2019dse,EventHorizonTelescope:2022wkp,Genzel:2010zy}. However, the singularity theorems \cite{Penrose:1964wq,Hawking:1970zqf,Senovilla:1998oua} indicate that under certain energy conditions, gravitational collapse inevitably leads to spacetime singularities, causing the failure of causal structure in those regions. To ensure spacetime completeness, one approach is the cosmic censorship conjecture \cite{Penrose:1969pc}, which proposes that singularities should be screened by event horizons, thereby protecting the causal structure of external spacetime. Another approach involves incorporating quantum gravity effects or modifying gravitational theory in the strong-field regime to suppress singularity formation, thus catalyzing the development of regular black hole models.

The regular black hole solutions are mainly obtained through two approaches: first, by introducing specific matter source terms in Einstein's field equations to regularize the singularity. This research direction began with the pioneering regular black hole model proposed by Bardeen  \cite{1968qtr..conf...87B}, subsequently expanded and deepened by Hayward \cite{Hayward:2005gi} and Ayon-Beato and Garcia  \cite{Ayon-Beato:1998hmi} based on different physical mechanisms, with other significant contributions in this direction seen in \cite{Dymnikova:1992ux,Bronnikov:2005gm,Fan:2016hvf,Lobo:2020ffi,Xu:2021lff,Ovalle:2023ref}, and other references;
The second approach is to eliminate singularities by modifying the gravitational theory framework itself, which includes two main branches: (1) modifying classical gravitational theories, such as f(R) gravity, Lovelock gravity, etc. (see, e.g.,  \cite{Bueno:2024dgm, Junior:2023qaq, Bakopoulos:2023fmv, Fabris:2023opv,Estrada:2024uuu}); (2) introducing quantum effects at the semi-classical or quantum gravity level to avoid singularity formation through quantum corrections  (see, e.g.,  \cite{Boehmer:2007ket,Casadio:2023iqt,Nicolini:2019irw,Knorr:2022kqp,Brannlund:2008iw,Lewandowski:2022zce,Bonanno:2023rzk,Akil:2022coa,Bonanno:2000ep,Pawlowski:2023dda}).

In quantum gravity theory, asymptotic safety gravity is characterized by its gravitational coupling constant approaching a non-trivial Reuter fixed point in the high-energy limit, providing a theoretically self-consistent and ultraviolet-complete framework that naturally realizes black hole singularity regularization. Recently, Bonanno et al.  \cite{Bonanno:2023rzk} extended the ideas of Markov and Mukhanov \cite{Markov:1985py} by constructing an effective Lagrangian to obtain a novel regular black hole model. This model regulates the metric through a scale parameter $\xi$, which cannot be determined from first principles and can only be constrained through observations \cite{Bonanno:2023rzk}.
Although the traditional quantum mechanical viewpoint considers $\xi$ as a quantum effect scale parameter that primarily operates near the Planck scale, with negligible influence on the black hole exterior, extensive studies on Hawking radiation and black hole evolution (relevant literature includes \cite{Hawking:1975vcx, Giddings:2017jts, Giddings:2019ujs, Maldacena:2013xja, Giddings:2009ae, Haggard:2016ibp, Almheiri:2012rt}, etc.) demonstrate that quantum effects exist near black hole event horizons and influence the event horizon scale, breaking this traditional understanding. Furthermore, within the asymptotic safety gravity framework, quantum effects are manifested through the scale dependence of Newton's coupling constant. This unique mechanism, in principle, allows the quantum effect scale to deviate significantly from the Planck scale \cite{Held:2019xde}, thereby enabling observable effects at larger scales. However, these quantum effect influences are primarily concentrated near black hole event horizons, and in regions far from the event horizon, quantum effects decay and degrade to classical black hole spacetime structure. Given that the $\xi$ parameter produces significant effects only in strong-field regions, observational constraints on it require extremely high-precision gravitational field detection capabilities. Future space-based gravitational wave detectors, through measuring the fine structural characteristics of gravitational wave signals from extreme mass ratio inspiral systems, are expected to achieve effective constraints on this theoretical parameter, thereby testing the physical feasibility of regular black hole models under asymptotic safety gravity.

On the other hand, gravitational wave signals generated by EMRI systems formed by stellar-mass compact objects (CO) orbiting supermassive black holes at galactic centers will become one of the most promising sources for future space-based detectors (such as the Laser Interferometer Space Antenna (LISA) \cite{Danzmann:1997hm,LISA:2017pwj},  Taiji \cite{Hu:2017mde} and TianQin \cite{TianQin:2015yph}). 
Through precise detection of these gravitational waveforms, not only can a wealth of information about black hole spacetime structure be obtained and physical parameters measured more accurately (see e.g., \cite{Babak:2017tow,Berry:2019wgg,Fan:2020zhy,Zi:2021pdp}), but it also provides a real experimental platform for testing general relativity (see e.g., \cite{Shen:2025svs,Kumar:2024dql,Kumar:2024utz,Destounis:2021mqv}). Furthermore, EMRIs have made significant progress in testing the no-hair theorem and constraining parameter ranges (see e.g., \cite{Ryan:1997hg,Barack:2006pq,Chua:2018yng,Zi:2025idv,DellaRocca:2024sda,Gair:2012nm,Zi:2021pdp,Kumar:2024our,Datta:2019euh,Meng:2024cnq,Zhao:2024exh,Niu:2019ywx,Zhang:2024csc,Tan:2024utr,Jiang:2021htl}). 
At the same time, through the analysis of modulation effects induced by matter distribution around supermassive black holes on gravitational wave waveforms, novel pathways for probing black hole environments (including dark matter) are provided
(see e.g., \cite{Barausse:2006vt,Barausse:2007dy,Yunes:2011ws,Barausse:2014tra,Zhang:2024ogc,Zhang:2024hrq,Hannuksela:2019vip,Zhang:2024ugv,Cardoso:2022whc,Duque:2023seg,Rahman:2023sof,Destounis:2022obl}). 
In particular, EMRIs possess unique advantages in detecting quantum gravitational effects. Although quantum gravitational effects are extremely small within a single orbital period during the inspiral phase, the unique feature of EMRIs lies in the fact that small-mass objects typically undergo thousands to tens of thousands of orbital cycles during their inspiral process in the strong-field region. Through long-term accumulation during the inspiral phase, tiny quantum effects may be amplified through phase accumulation, ultimately leaving observable characteristic signatures in gravitational wave waveforms, which could potentially be detected by future space-based gravitational wave detectors.
Related research on detecting quantum effects through EMRIs has been conducted in multiple studies (see e.g., \cite{Tu:2023xab,Liu:2024qci,Yang:2024lmj,Yang:2024cnd,Fu:2024cfk,Zi:2024jla,Zi:2023qfk}). This detection precision far exceeds the range achievable by other weak-field gravitational experiments, making EMRIs an important astrophysical platform for detecting quantum gravitational effects.

Based on the high-precision detection capability of EMRIs for background spacetime structure, this study adopts the regular black hole model proposed by Bonanno et al. under the asymptotically safe gravity framework \cite{Bonanno:2023rzk} as the central supermassive black hole in EMRIs. The spacetime geometry of this black hole model is dominated by the scale parameter $\xi$, whose value cannot be derived from first principles and can only be constrained through observational data \cite{Bonanno:2023rzk}. Discussions about $\xi$ have been initiated in multiple domains, including quasi-normal modes \cite{Stashko:2024wuq}, strong gravitational lensing effects \cite{Gao:2024cgg}, and black hole shadow observations \cite{Sanchez:2024sdm}. In this paper, we focus on analyzing the leading-order effects of the parameter, first deriving the geodesic orbital frequencies, energy flux, angular momentum flux, and orbital evolution equations for a non-spinning stellar-mass compact object in the equatorial plane. Subsequently, we employ the AAK method to construct complete gravitational wave waveform templates. By calculating the phase accumulation difference relative to a Schwarzschild black hole and combining the noise characteristics of the LISA detector, we apply mismatch and Fisher information matrix analysis methods to quantitatively evaluate LISA's measurement precision and physical constraint capabilities for $\xi$.

The structure of this paper is as follows: Section \ref{sec2} introduces the black hole background and geodesics, Section \ref{sec3} elucidates the gravitational wave construction method, including flux calculations, orbital evolution, and waveform analysis methods; Section \ref{sec4} presents the main research results and discussion; Section \ref{sec5} summarizes the entire paper. Except for the data processing section, this paper adopts geometric units, i.e., c = G = 1.

\section{\label{sec2}Regular Black Hole Background and Geodesics in Asymptotically Safe Gravity}
\subsection{Background Metric}

Bonanno et al. extended the ideas of Markov and Mukhanov \cite{Markov:1985py} by introducing an effective Lagrangian within the asymptotically safe gravity framework to study dust collapse processes, ultimately obtaining a  regular black hole model \cite{Bonanno:2023rzk}. Specifically, they adopted the system action \cite{Markov:1985py}
\begin{equation}
S=\frac{1}{16\pi G_N}\int{d^4x}\sqrt{-g}\left[R+2\chi\left(\epsilon\right)\mathcal{L}\right],
\label{1}
\end{equation}
where $\mathcal{L}=\ -\epsilon$ is the matter Lagrangian, $\epsilon$ is the fluid density, and $\chi\left(\epsilon\right)$ is a coupling function reflecting how the gravitational constant varies with energy density. Within this framework, the gravitational field equations are modified to
\begin{equation}
R_{\mu\nu}-\frac{1}{2}g_{\mu\nu}R=8\pi G\left(\epsilon\right)T_{\mu\nu}-\mathrm{\Lambda}\left(\epsilon\right)g_{\mu\nu},
\label{2}
\end{equation}
where $T_{\mu\nu}=\left(\epsilon+p(\epsilon)\right) u_\mu u_\nu+pg_{\mu\nu}$ is the energy-momentum tensor, with the effective gravitational constant $G\left(\epsilon\right)$ and cosmological constant $\Lambda\left(\epsilon\right)$ given by \cite{Bonanno:2023rzk}
\begin{equation}
8\pi G\left(\epsilon\right)=\frac{\partial\left(\chi\epsilon\right)}{\partial\epsilon},\quad\Lambda\left(\epsilon\right)=-\epsilon^2\frac{\partial\chi}{\partial\epsilon}.
\label{3}
\end{equation}

In the theoretical framework of asymptotically safe  gravity, the gravitational constant is assumed to be dominated by renormalization group trajectories approaching the ultraviolet (UV) fixed point \cite{Reuter:1996cp,Bonanno:2019ilz,Bonanno:2021squ}. Under this assumption, the gravitational constant $G\left(\epsilon\right)$ can be expressed as
\begin{equation}
G\left(\epsilon\right)=\frac{G_N}{1+\xi\epsilon},
\label{4}
\end{equation}
where $\xi$ represents the scale parameter, describing the intensity of quantum gravity corrections, whose specific magnitude needs to be constrained through observations \cite{Bonanno:2023rzk}.

During the process of gravitational collapse of dust (assuming a pressureless fluid with $p = 0$), asymptotically safe gravity theory predicts a regular spherically symmetric black hole solution \cite{Bonanno:2023rzk}, with metric
\begin{equation}
ds^2=-f(r)dt^2+f\left(r\right)^{-1}dr^2+r^2d\Omega^2,
\label{5}
\end{equation}
where
\begin{equation}
f\left(r\right)=1-\frac{r^2}{3\xi}\ln{\left(1+\frac{6M\xi}{r^3}\right)}.
\label{6}
\end{equation}

Evidently, in the limit of $r\rightarrow 0$, this metric maintains regularity, effectively avoiding the pathological behavior caused by singularities in traditional solutions; in the limit of scale parameter $\xi\rightarrow 0$ (or $r\rightarrow\infty$), this solution degenerates to the classical Schwarzschild black hole solution. It can be seen that asymptotically safe gravity effectively regulates gravitational behavior at high energy scales through renormalization group trajectories, thereby constructing singularity-free spacetime geometry within the theoretical framework. This characteristic demonstrates the theoretical advantages of asymptotically safe gravity theory in its self-consistency at ultra-high energy scales and its ability to avoid singularity problems.

\begin{figure*}[]
\includegraphics[width=1 \textwidth]{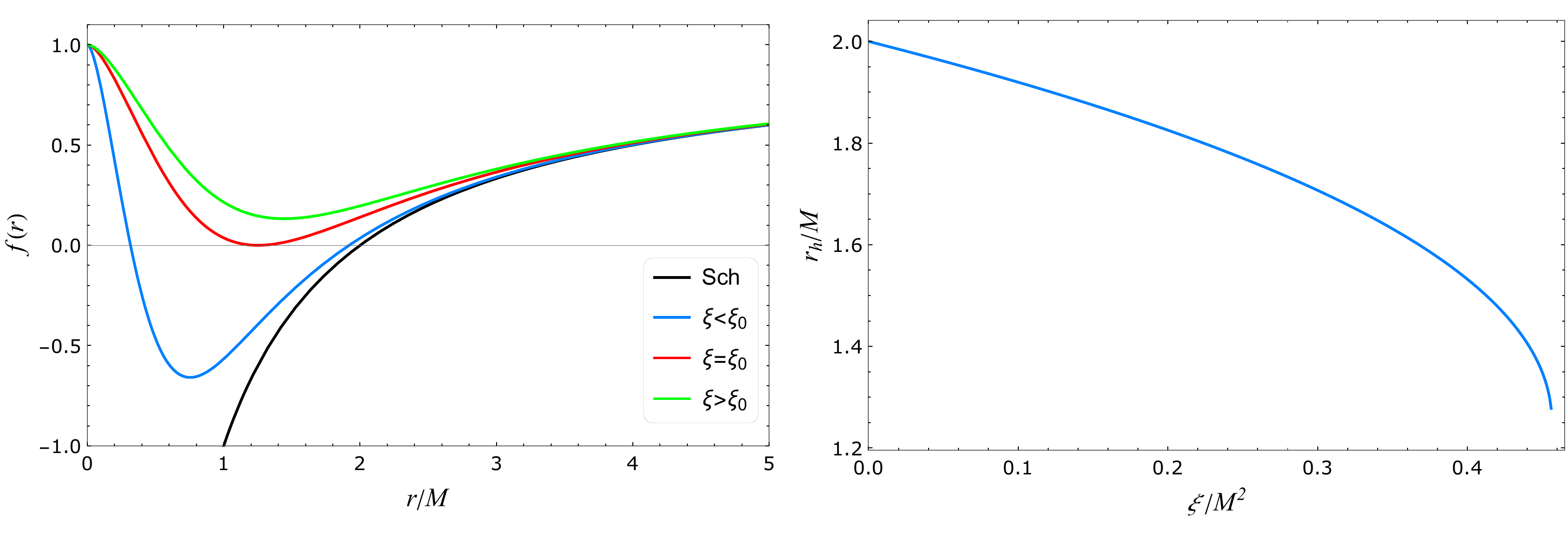}
\caption{
The left figure shows the existence conditions for the event horizon of a regular black hole in asymptotically safe gravity, where $\frac{\xi_0}{M^2}\approx0.4565$ represents an extremal black hole; the right figure illustrates the effect of the scale parameter $\xi$ on the outer event horizon radius.}
\label{a}
\end{figure*}

Fig. \ref{a} illustrates the event horizon structure of a regular black hole in asymptotically safe gravity. The left figure shows that: when $\xi<\xi_0$, metric (\ref{6}) has two event horizons (inner and outer horizons); when $\xi=\xi_0$, the inner and outer horizons coincide, corresponding to an extremal black hole configuration; when $\xi>\xi_0$, metric (\ref{6}) exhibits a horizon-free compact geometric structure. Notably, in regions far from the event horizon, this metric is asymptotically consistent with the Schwarzschild solution, indicating that conventional weak-field tests struggle to break this degeneracy, and detecting the effects of scale parameter $\xi$ requires investigation in strong gravitational field regions. The right figure shows that compared to a Schwarzschild black hole, the event horizon radius of a regular black hole in asymptotically safe gravity is significantly smaller than that of a Schwarzschild black hole due to the influence of scale parameter $\xi$ (when scale parameter $\xi\rightarrow 0$, a regular black hole in asymptotically safe gravity degenerates into a Schwarzschild black hole), a phenomenon that is a common characteristic of many regular black hole models.

\subsection{Timelike Geodesics}
In an extreme mass ratio (EMR) system, a stellar-mass CO orbiting a supermassive black hole (a regular black hole in asymptotically safe gravity) follows timelike geodesics. When the stellar-mass CO is restricted to move in the equatorial plane, i.e., $\theta=\pi/2$, the Lagrangian for the regular black hole in asymptotically safe gravity (\ref{6}) is
\begin{equation}
\mathcal{L}=\frac{1}{2}mg_{\mu\nu}\frac{{\rm dx}^\mu}{d\tau}\frac{{\rm dx}^\nu}{d\tau}=\frac{1}{2}m\left(-f\left(r\right)\dot{t^2}+\frac{1}{f\left(r\right)}\dot{r^2}+r^2\dot{\phi^2}\right),
\label{7}
\end{equation}
where the dot represents differentiation with respect to proper time $\tau$. Since the black hole metric we consider is static and spherically symmetric, there exist two conserved quantities, namely the energy $E$ and angular momentum $L$. At this time, the equations of motion of the system can be derived from the Lagrangian (\ref{7}) as
\begin{equation}
\frac{dt}{d\tau}=\frac{E}{mf\left(r\right)},
\label{8}
\end{equation}
\begin{equation}
\left(\frac{dr}{d\tau}\right)^2=\frac{E^2}{m^2}-f\left(r\right)\left(1+\frac{L^2}{m^2r^2}\right),
\label{9}
\end{equation}
\begin{equation}
\frac{d\phi}{d\tau}=\frac{L}{mr^2},
\label{10}
\end{equation}
\begin{equation}
\frac{d\theta}{d\tau}=0.
\label{11}
\end{equation}

To more effectively describe the eccentric orbit of the stellar-mass CO, we parameterize the radial coordinate $r$ with the parametric equation
\begin{equation}
r\left(\psi\right)=\frac{Mp}{1+e\ cos\psi},
\label{12}
\end{equation}
where $p$ is the dimensionless semi-latus rectum and $e$ is the orbital eccentricity. When the orbit sweeps through a complete cycle, the parameter $\psi$ varies from 0 to $2\pi$. The two turning points of the radial motion of the orbit can be expressed as
\begin{equation}
r_p=\frac{Mp}{1+e},\quad r_a=\frac{Mp}{1-e}.
\label{13}
\end{equation}
At these two turning points, the radial velocity of the orbit is zero, i.e., $\frac{dr}{d\tau}=0$. Combining equations (\ref{9}) and (\ref{13}), the expressions for energy and angular momentum can be derived as
\begin{align}
E^2=\frac{4 e m^2 \left(3 (e-1)^2 \xi -p^2 \ln \left(1-\frac{6 (e-1)^3 \xi }{p^3}\right)\right) \left(3 (e+1)^2 \xi -p^2 \ln
   \left(\frac{6 (e+1)^3 \xi }{p^3}+1\right)\right)}{3 \left(e^2-1\right)^2 \xi  \left(12 e \xi +p^2 \ln \left(1-\frac{6
   (e-1)^3 \xi }{p^3}\right)-p^2 \ln \left(\frac{6 (e+1)^3 \xi }{p^3}+1\right)\right)},
\label{14}
\end{align}

\begin{align}
L^2=-\frac{m^2 M^2 p^4 \left((e+1)^2 \ln \left(1-\frac{6 (e-1)^3 \xi }{p^3}\right)-(e-1)^2 \ln \left(\frac{6 (e+1)^3 \xi
   }{p^3}+1\right)\right)}{\left(e^2-1\right)^2 \left(12 e \xi +p^2 \ln \left(1-\frac{6 (e-1)^3 \xi }{p^3}\right)-p^2 \ln
   \left(\frac{6 (e+1)^3 \xi }{p^3}+1\right)\right)}.
\label{15}
\end{align}
Here, the scale parameter $\xi$ has undergone a dimensionless transformation, i.e., $\xi\rightarrow M^2\xi$. 
It should be particularly noted that in the subsequent theoretical derivations and numerical calculations in this work, all scale parameters $\xi$ involved refer to dimensionless parameters unless otherwise explicitly specified.
Obviously, when the scale parameter $\xi$ vanishes, expressions (\ref{14}) and (\ref{15}) revert to the results for a Schwarzschild black hole \cite{Hopper:2015icj,Cutler:1994pb}. 

Since the motion of the stellar-mass CO under consideration is restricted to the equatorial plane, there exist only two fundamental frequencies: the radial frequency $\Omega_r$ and the azimuthal frequency $\Omega_\phi$. They are associated with radial and azimuthal motions, respectively, and can be expressed as
\begin{equation}
\Omega_r=\frac{2\pi}{T_r},\quad T_r=\int_{0}^{t_0}{dt=\int_{0}^{2\pi}{\frac{dt}{d\psi}d\psi}},
\label{16}
\end{equation}
\begin{equation}
\Omega_\phi=\frac{\Delta\phi}{T_r},\quad \Delta\phi=\int_{0}^{\phi_0}{d\phi=\int_{0}^{2\pi}{\frac{d\phi}{d\psi}d\psi}}.
\label{17}
\end{equation}
Where $T_r$ is the radial period and $\Delta\phi$ is the azimuthal displacement. By combining equations (\ref{8}) and (\ref{10}) and expanding, we can obtain
\begin{align}
\Omega_r=&\frac{X^3}{Mp^{3/2}}-\frac{3X^5}{Mp^{5/2}}-\frac{3X^5\left(6e^2+5X-2\right)}{2Mp^{7/2}}
-\frac{3X^5\left[18e^4+2(15X-8)e^2-2\left(X+3\right)\xi-5X+14\right]}{2Mp^{9/2}}\nonumber\\
&+O\left(p^{-11/2}\right),
\label{18}
\end{align}
\begin{align}
\Omega_\phi=&\frac{X^3}{Mp^{3/2}}+\frac{3e^2X^3}{Mp^{5/2}}+\frac{3X^3\left[12e^4+\left(10X-3\right)e^2-10\left(X-1\right)\right]}{4Mp^{7/2}}
\nonumber\\
&+\frac{3X^3\left\{36e^6+(60X-29)e^4-2\left[\left(2X+9\right)\xi+20X-39\right]e^2+4\left[\left(X-3\right)\xi-5X+5\right]\right\}}{4Mp^{9/2}}\nonumber\\&+O\left(p^{-11/2}\right).
\label{19}
\end{align}
Where $X=\sqrt{1-e^2}$. From expressions (\ref{14}) and (\ref{15}), it can be seen that the influence of the scale parameter $\xi$ is primarily manifested in higher-order expansion terms. When $\xi=0$, our results revert to the case of a Schwarzschild black hole.

\section{\label{sec3}Method}
\subsection{ Flux and Orbital Evolution}
In the previous section, we obtained the fundamental frequencies ($\Omega_r$ and $\Omega_\phi$) on the equatorial plane through geodesic analysis, but these results did not consider the effects of gravitational radiation. This section will focus on deriving orbital evolution in the regular black hole background, which is primarily driven by energy flux and angular momentum flux caused by gravitational radiation. For energy flux and angular momentum flux, we adopt expressions derived by Peters and Mathews using the standard quadrupole approximation  for calculation \cite{Peters:1964zz,Peters:1963ux},and their expressions are as follows
\begin{equation}
\left\langle\frac{dE}{dt}\right\rangle=\frac{1}{5}\mu\left\langle\frac{d^3Q_{ij}}{dt^3}\frac{d^3Q^{ij}}{dt^3}-\frac{1}{3}\frac{d^3Q_{ii}}{dt^3}\frac{d^3Q^{jj}}{dt^3}\right\rangle,
\label{20}
\end{equation}
\begin{equation}
\left\langle\frac{dL_i}{dt}\right\rangle=\frac{2}{5\mu M}\epsilon_{ijk}\left\langle \frac{d^2Q_{jm}}{dt^2}\frac{d^3Q^{km}}{dt^3}\right\rangle.
\label{21}
\end{equation}
Here, $\mu=mM/(m+M)$ is the reduced mass, which in extreme mass ratio inspiral systems can be approximated as the stellar-mass compact object, i.e., $\mu\approx m$. $Q_{ij}$ is the inertia tensor, expressed as
\begin{equation}
Q_{ij}=\mu x^i x^j,
\label{22}
\end{equation}
where $x^i$ is the position vector between the stellar-mass CO and the regular black hole in asymptotically safe gravity, represented in spherical coordinates on the equatorial plane as $x^i=(r\cos\phi,r\sin\phi,0)$. 
In the weak-field approximation, the average energy flux and angular momentum flux can be expanded into expressions containing scale parameters. The specific forms are
\begin{equation}
\left\langle\frac{dE}{dt}\right\rangle=\left\langle\frac{dE}{dt}\right\rangle_{GR}+\frac{F(\xi,e)\mu^2}{M^2p^{8}},
\label{23}
\end{equation}
\begin{equation}
\left\langle\frac{dL}{dt}\right\rangle=\left\langle\frac{dL}{dt}\right\rangle_{GR}+\frac{G(\xi,e)\mu^2}{M p^{13/2}}.
\label{24}
\end{equation}
The functions $F\left(\xi,e\right)$ and $G\left(\xi,e\right)$ represent higher-order correction terms carrying the scale parameter $\xi$, which appear at the 3PN order in the Post-Newtonian (PN) expansion. Since the expressions of these functions are rather complex and lengthy, the detailed PN expansion content and their specific forms are given in Appendix \ref{appA}. From the structure of the equations, it can be seen that the scale parameter $\xi$ does not directly affect the leading-order contribution of the average flux, but rather manifests its influence on gravitational wave radiation through higher-order correction terms. It should be particularly noted that the correction terms $F\left(\xi,e\right)$ and $G\left(\xi,e\right)$ characterize the deviation effects of the scale parameter $\xi$ relative to General Relativity. These correction terms naturally reduce to the pure General Relativity 3PN order contributions in the limit $\xi\rightarrow0$.

Orbital evolution is primarily driven by energy and angular momentum dissipation due to gravitational wave radiation. Under the adiabatic approximation, changes in the system's orbital energy and angular momentum are completely converted into energy radiated as gravitational waves, thus expressed as
\begin{equation}
{\dot{E}}_{\mathrm{GW}}=-\left\langle\frac{dE}{dt}\right\rangle=-\mu\dot{E},
\label{25}
\end{equation}
\begin{equation}
{\dot{L}}_{\mathrm{GW}}=-\left\langle\frac{dL}{dt}\right\rangle=-\mu\dot{L}.
\label{26}
\end{equation}
Since $E$ and $L$ in expressions (\ref{25}) and (\ref{26}) are functions of $p$ and $e$, chain rule differentiation yields
\begin{equation}
-{\dot{E}}_{\mathrm{GW}}=\mu\frac{\partial E}{\partial p}\frac{dp}{dt}+\mu\frac{\partial E}{\partial e}\frac{de}{dt},
\label{27}
\end{equation}
\begin{equation}
-{\dot{L}}_{\mathrm{GW}}=\mu\frac{\partial L}{\partial p}\frac{dp}{dt}+\mu\frac{\partial L}{\partial e}\frac{de}{dt}.
\label{28}
\end{equation}
Combining expressions (\ref{27}) and (\ref{28}), we can obtain the orbital evolution equations ($\dot{p}\left(t\right)$ and $\dot{e}\left(t\right)$) for a stellar-mass CO in a regular black hole in asymptotically safe gravity, expressed as
\begin{equation}
m\frac{dp}{dt}=\left[\frac{\partial E}{\partial e}{\dot{L}}_{\mathrm{GW}}-\frac{\partial L}{\partial e}{\dot{E}}_{\mathrm{GW}}\right]/\left[\frac{\partial E}{\partial p}\frac{\partial L}{\partial e}-\frac{\partial E}{\partial e}\frac{\partial L}{\partial p}\right],
\label{29}
\end{equation}
\begin{equation}
m\frac{de}{dt}=\left[\frac{\partial L}{\partial p}{\dot{E}}_{\mathrm{GW}}-\frac{\partial E}{\partial p}{\dot{L}}_{\mathrm{GW}}\right]/\left[\frac{\partial E}{\partial p}\frac{\partial L}{\partial e}-\frac{\partial E}{\partial e}\frac{\partial L}{\partial p}\right].
\label{30}
\end{equation}

\begin{figure*}[]
\includegraphics[width=1 \textwidth]{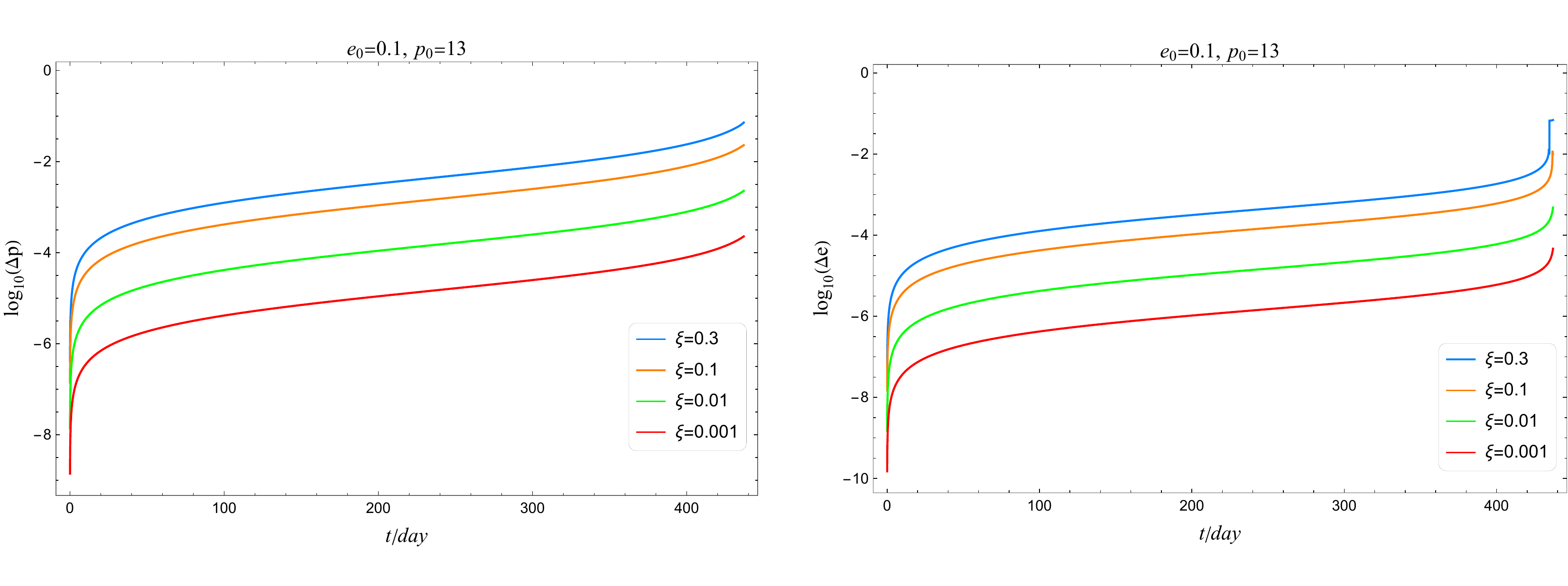}
\caption{
The deviation $\Delta X=X_{\xi\ne 0}-X_{\xi= 0}$ in the semi-latus rectum $p$ and eccentricity $e$ over time evolution under different scale parameters $\xi$, where $X=\{p,e\}$. The corresponding initial conditions are $p_0=13, e_0=0.1$.}
\label{b}
\end{figure*}

Substituting expressions (\ref{14}), (\ref{15}), (\ref{23}), and (\ref{24}) into (\ref{29}) and (\ref{30}) yields the orbital evolution equations for the stellar-mass CO. Notably, orbital evolution ($\dot{p}\left(t\right)$ and $\dot{e}\left(t\right)$) will be influenced by the scale parameter $\xi$. Although this influence does not appear in the leading-order terms of the orbital parameter $p$, it regulates the evolution of higher-order terms. To intuitively present the impact of the scale parameter on orbital evolution, we analyze it by calculating the orbital evolution deviation relative to a Schwarzschild black hole, with results shown in Fig. \ref{b}. Evidently, as time progresses, the orbital evolution deviations $\Delta p, \Delta e$ between a regular black hole in asymptotically safe gravity and a Schwarzschild black hole gradually increase, reaching maximum before the stellar-mass CO plunges into the black hole. This indicates that the influence of scale parameter $\xi$ on orbital evolution accumulates over time, thereby providing a possibility for observational detection.

Furthermore, we can also calculate the time difference $\Delta t$ for a small body to reach the last stable orbit between the Schwarzschild black hole and the asymptotically safe gravity regular black hole cases. Considering the inspiral process starting from the initial semi-latus rectum $p_{in}$, the last stable orbit is reached when the orbital parameter evolves to $p = p_{LSO}$. By performing time integration on the orbital evolution equation (\ref{29})
\begin{equation}
t_{total} = \int_{p_{in}}^{p_{LSO}} \frac{dt}{dp} dp.
\label{100}
\end{equation}
Computing the inspiral times for the Schwarzschild black hole ($\xi = 0$) and the asymptotically safe gravity regular black hole ($\xi \neq 0$) cases respectively, we obtain the time difference between them
\begin{equation}
\Delta t = t_{total}^{\xi \neq 0} - t_{total}^{\xi = 0} = f\left(e, p_{in}, p_{LSO}, \xi\right).
\label{99}
\end{equation}
Here, the expression for $f\left(e, p_{in}, p_{LSO}, \xi\right)$ is extremely lengthy, but can be obtained through analytical integration of the orbital evolution equation.

Obviously, different scale parameters $\xi$ correspond to different last stable orbit positions, which affects the comparison of evolutionary timescales. In this study, we mainly ensure the reasonableness of comparison by appropriately adjusting the initial orbital parameters. Specifically, we set the initial condition $p_0 = 13$ along with different eccentricity values, such that the gravitational wave waveforms can evolve stably within a one-year observation time window for both Schwarzschild black holes and asymptotically safe gravity regular black holes, while always remaining far from their respective critical orbital regions. This parameter choice allows us to focus on analyzing the influence of the scale parameter $\xi$ on waveform characteristics. 

It should be noted that since we adopt a parameter configuration strategy that ensures orbital evolution always remains far from the respective innermost stable circular orbit (ISCO), the selected orbital parameters necessarily correspond to elliptical orbits with low eccentricity rather than circular orbits. Within this parameter domain, the orbital system exhibits low-eccentricity characteristics, maintaining moderate ellipticity throughout the entire evolution process without reaching complete circularization, which constitutes the typical orbital characteristics of EMRIs far from the ISCO region. Based on the above physical considerations, the following numerical evolution focuses on the gravitational wave radiation properties of low-eccentricity elliptical orbits, rather than the simplified case of idealized circular orbits.

Furthermore, as the orbit evolves, the phases corresponding to the two fundamental frequencies ($\Omega_r, \Omega_\phi$) also change, and their phase evolutions $\dot{\Phi_r}, \dot{\Phi_\phi}$ are expressed as
\begin{equation}
\frac{d\Phi_i}{dt}=\left\langle\Omega_i(p(t),e(t))\right\rangle,
\label{31}
\end{equation}
where $i=(r,\phi)$. To preliminarily assess the impact of scale parameter $\xi$ on EMRIs, we similarly employ phase deviation $\Delta\Phi_i$ at the same moment to quantify the effects of the scale parameter. The deviation is defined as $\Delta\Phi_i=\left|\Phi_i(\xi\ne 0)-\Phi_i(\xi= 0)\right|$.

It should be emphasized that the phase difference $\Delta\Phi_i$ defined here is an intermediate physical quantity used for theoretical analysis, primarily employed to analyze the influence of scaling parameters on gravitational wave phases, rather than being an observable directly measured by gravitational wave detectors. In practice, gravitational wave detectors (such as LISA) receive complete time-domain gravitational wave signals containing both amplitude and phase modulation information. To construct the corresponding theoretical gravitational wave signals, appropriate waveform construction methods must be employed.

\subsection{Waveform Construction}

In this section, we employ the AAK method to construct gravitational waveforms excited by EMRIs. As an improved version of the Analytic Kludge (AK) method, the AAK method combines the computational efficiency of the AK method with the high-precision characteristics of the Numerical Kludge (NK) method, capable of significantly enhancing waveform accuracy while maintaining relatively low computational costs. As described in the literature \cite{Chua:2017ujo}, the AAK method can effectively increase the number of detectable EMRI events compared to the traditional AK method; while compared to the NK method, the AAK method improves computational speed by an order of magnitude. For detailed content regarding these methods, please refer to the literature \cite{Barack:2003fp,Babak:2006uv,Chua:2015mua,Chua:2017ujo}.

To construct the corresponding waveforms, we adopt a coordinate system adapted to the detector, where the unit vector $\hat{r}$ points toward the source position, and the other two unit vectors $\hat{p}$ and $\hat{q}$ are defined as \cite{Barack:2003fp}
\begin{equation}
\hat{p}=\frac{\hat{r}\times\hat{L}}{\left|\hat{r}\times\hat{L}\right|}, \quad \hat{q}=\hat{p}\times\hat{r}.
\label{32}
\end{equation}
In the transverse-traceless gauge, the waveform is represented via the quadrupole approximation as
\begin{equation}
h_{ij}=\frac{2}{D_L}\left(P_{ik}P_{jl}-\frac{1}{2}P_{ij}P_{kl}\right){\ddot{I}}^{kl}, \; h^{\left(+,\times\right)}=\frac{1}{2}h_{ij}H_{ij}^{\left(+,\times\right)}.
\label{33}
\end{equation}
Where $D_L$ is the luminosity distance.The projection tensor $P_{ij}$ is defined as $P_{ij}=\delta_{ij}-{\hat{r}}_i{\hat{r}}_j$, and the polarization basis tensors $H_{ij}^+$ and $H_{ij}^\times$ are defined as
\begin{equation}
H_{ij}^+=p_ip_j-q_iq_j, \quad H_{ij}^\times=p_iq_j+q_ip_j.
\label{34}
\end{equation}
The gravitational waveform can be represented as a superposition of $n$-th harmonic components
\begin{equation}
h^+=A\sum_{n} h_n^+, \quad h^\times=A\sum_{n} h_n^\times,
\label{35}
\end{equation}
where $A=\frac{(M\Omega_\phi)^{2/3}m}{D_L}$, and the expressions for $h_n^+$ and $h_n^\times$ are \cite{Barack:2003fp}
\begin{align}
h_n^+=\left[1+\left(\hat{r}\cdot\hat{L}\right)^2\right]\left[b_n\sin{\left(2\gamma\right)}-a_n\cos{\left(2\gamma\right)}\right]
+\left[1-\left(\hat{r}\cdot\hat{L}\right)^2\right]c_n,
\label{36}
\end{align}
\begin{equation}
h_n^\times=2\left(\hat{r}\cdot\hat{L}\right)\left[b_n\cos{\left(2\gamma\right)}+a_n\sin{\left(2\gamma\right)}\right].
\label{37}
\end{equation}
Where $\gamma=\Phi_\phi-\Phi_r$ represents the azimuthal angle, and coefficients $a_n$, $b_n$, and $c_n$ can be expressed through Bessel functions of the first kind \cite{Barack:2003fp}.

Within the LISA detector framework, the two polarized gravitational wave signals are transformed into detector measurement outputs via specific response functions, specifically given as
\begin{equation}
h^{\mathrm{I,II}}\left(t\right)=\frac{\sqrt3}{2}\left[h_+\left(t\right)F_+^{\mathrm{I,II}}\left(t\right)+h_\times\left(t\right)F_\times^{\mathrm{I,II}}\left(t\right)\right].
\label{38}
\end{equation}
Where $F_+^{\mathrm{I,II}}\left(t\right)$ and $F_\times^{\mathrm{I,II}}\left(t\right)$ are the LISA antenna pattern functions, which depend on the geometric configuration of the source direction $\left(\theta_S,\phi_S\right)$ and black hole spin direction $\left(\theta_1,\phi_1\right)$ in the ecliptic coordinate system \cite{Barack:2003fp,Cutler:1997ta,Apostolatos:1994mx}.

\subsection{Mismatch and Fisher Information Matrix}
To quantify the effect of the scale parameter $\xi$ on the waveform, we compare the overlap between the regular black hole waveform signal with $\xi \neq 0$ and the traditional black hole waveform with $\xi = 0$. The overlap $\mathcal{O}\left(h_a\middle| h_b\right)$ is defined as
\begin{equation}
\mathcal{O}\left(h_a\middle| h_b\right)=\frac{\left\langle h_a\middle| h_b\right\rangle}{\sqrt{\left\langle h_a\middle| h_a\right\rangle\left\langle h_b\middle| h_b\right\rangle}}.
\label{39}
\end{equation}
Where the noise-weighted inner product $\left\langle h_a\middle| h_b\right\rangle$ is
\begin{equation}
\left\langle h_a\middle| h_b\right\rangle=2\int_{0}^{\infty}{df\frac{h_a^\ast\left(f\right)h_b\left(f\right)+h_a\left(f\right)h_b^\ast\left(f\right)}{S_n\left(f\right)}}.
\label{40}
\end{equation}
Here, $h_{a,b}\left(f\right)$ is the Fourier transform of the time-domain signal. $S_n$ is the noise power spectral density of the detector, such as LISA \cite{Maselli:2021men,Robson:2018ifk}, TianQin \cite{TianQin:2020hid}, and Taiji \cite{Ruan:2018tsw}.

Based on the definition of overlap, the mismatch $\mathcal{M}$ can be directly expressed as
\begin{equation}
\mathcal{M}=1-\mathcal{O}\left(h_a\mid h_b\right).
\label{41}
\end{equation}
When two signals are identical, the overlap is $\mathcal{O}=1$ and the mismatch is $\mathcal{M}=0$. As described in the literature \cite{Flanagan:1997kp,Lindblom:2008cm,Zi:2023qfk}, to effectively distinguish between two waveform signals, the mismatch must satisfy $\mathcal{M}\geq D/2\rho^2$, where $\rho$ is the signal-to-noise ratio and $D$ represents the number of intrinsic parameters in the system. 
In this paper, there are only six intrinsic parameters. For EMRIs, the LISA detection threshold is typically set at a signal-to-noise ratio of 20 \cite{Babak:2017tow}. Therefore, the mismatch threshold for distinguishing gravitational wave signals from regular black holes in asymptotically safe gravity is \(\mathcal{M}=0.0075\).

The aforementioned mismatch analysis is based on linear approximation and does not adequately account for the correlations between various parameters in EMRIs, potentially introducing systematic errors. To more accurately assess LISA's capability for measuring scaling parameters, we further employ the Fisher information matrix method to analyze parameter estimation precision under higher signal-to-noise ratio conditions \cite{Vallisneri:2007ev}. The Fisher information matrix is defined as

\begin{equation}
\Gamma_{ij}=\left\langle\frac{\partial h}{\partial\lambda_i}\middle|\frac{\partial h}{\partial\lambda_j}\right\rangle,
\label{42}
\end{equation}
where $\lambda_i$ represents the set of relevant parameters that constitute the gravitational wave signal (\ref{38}). Under high SNR conditions, the variance-covariance matrix for parameter estimation can be approximated using the inverse of the FIM, namely
\begin{equation}
\Sigma_{ij}\equiv\left\langle\delta\lambda_i\delta\lambda_j\right\rangle=\left(\Gamma^{-1}\right)_{ij}.
\label{43}
\end{equation}
The statistical uncertainty for a specific parameter is given by the square root of the diagonal elements of the variance-covariance matrix (\ref{43})
\begin{equation}
\sigma_i=\Sigma_{ii}^{1/2}.
\label{44}
\end{equation}

The space-based gravitational wave detectors, represented by LISA, adopt a dual-channel interferometer design, achieving high-precision waveform reconstruction through joint measurement data from both channels. In this multi-channel framework, the calculation of signal-to-noise ratio and covariance matrix should consider the contributions from these two channels. Specifically, the total system signal-to-noise ratio can be expressed as \cite{Fu:2024cfk,Zhang:2024ogc}
\begin{equation}
\label{eq:total_snr}
\rho=\sqrt{\left\langle h_1\middle| h_1\right\rangle+\left\langle h_2\middle| h_2\right\rangle}.
\end{equation}
The diagonal elements of the covariance matrix are
\begin{equation}
\label{eq:diagonal_covariance}
{\sigma_i}^2={\left(\Gamma_1+\Gamma_2\right)_{ii}}^{-1}.
\end{equation}

\section{\label{sec4}Results}

In this section, to demonstrate the differences between gravitational wave signals from a regular black hole in asymptotically safe gravity and a traditional black hole, we set the parameter configuration: observation time of one year, $M=10^6M_\odot$, $m=10M_\odot$, $D_L=1$ Gpc, $\theta_s=\frac{\pi}{3}$, $\phi_s=\frac{\pi}{2}$, $\theta_1=\frac{\pi}{4}$, $\phi_1=\frac{\pi}{4}$. The initial orbital parameters $e_0$ and $p_0$ are marked in the corresponding figures. When selecting $p_0$, we ensure that the inspiral process during the observation period always remains far from the truncation point of the last stable orbit \cite{Stein:2019buj}, thereby guaranteeing that the waveform evolution can fully continue for one year. Based on these parameters, we performed numerical evolution of the gravitational wave signals and compared the differences in waveform characteristics between the regular black hole and the traditional black hole. To evaluate the detectability of the scale parameter, we applied mismatch analysis and the FIM method to quantify waveform differences and provide theoretical predictions for scale parameter estimation precision.

\subsection{Waveform and Mismatch}

\begin{figure*}[]
\includegraphics[width=1 \textwidth]{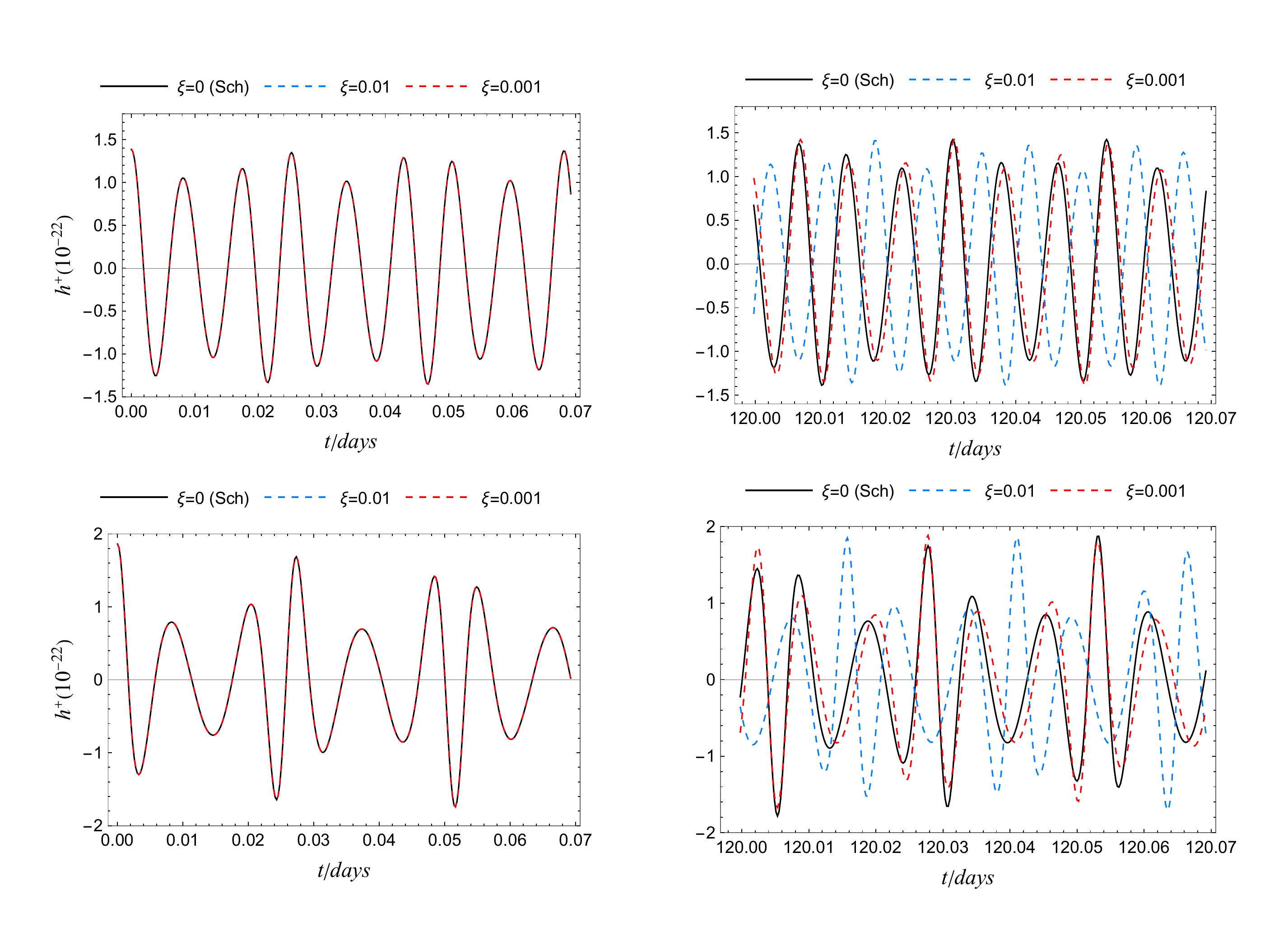}
\caption{
Initial conditions are set to $e_0=0.1$ (first row of figures), $e_0=0.3$ (second row of figures), and $p_0=13$, to study the effects of different scale parameters $\xi$ on gravitational waveforms. The black curves represent the gravitational waveforms of the Schwarzschild black hole ($\xi=0$), while the colored curves represent the gravitational waveforms of the regular black hole in the asymptotically safe gravity framework, with scale parameters $\xi=0.01$ (blue dashed lines) and $\xi=0.001$ (red dashed lines). The left column shows the waveforms at the early stage of evolution, and the right column shows the waveforms after $120$ days of evolution.}
\label{c}
\end{figure*}

Fig. \ref{c} presents the impact of the scale parameter $\xi$ on the gravitational waveform $h^+$, evolving under initial eccentricity conditions of $e_0=0.1$ and $e_0=0.3$. All waveforms begin their evolution from identical initial conditions, with the left column showing the waveform in the early evolution stage and the right column showing the waveform after 120 days of evolution. The black curves represent the Schwarzschild black hole ($\xi=0$) waveform, while the blue dashed lines and red dashed lines correspond to the regular black hole waveforms with scale parameters $\xi=0.01$ and $\xi=0.001$, respectively. The results indicate that during the early evolution stage, the waveforms under different scale parameters almost completely overlap. However, as time progresses, the correction effects introduced by the scale parameter in the regular black hole gradually accumulate,
 leading to significantly increased differences in the waveforms. The larger the value of scale parameter $\xi$, the more pronounced the deviation in the waveform.

\begin{figure*}[]
\includegraphics[width=1 \textwidth]{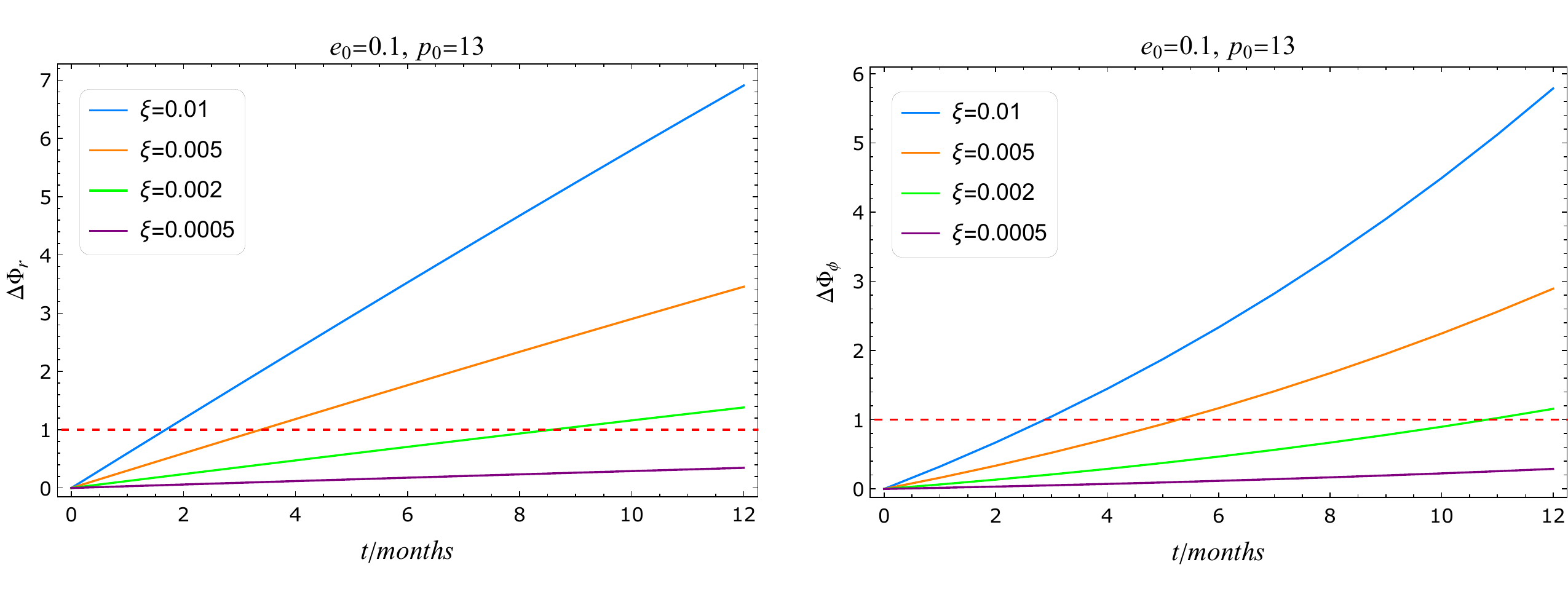}
\caption{
Under fixed conditions of $e_0=0.1$ and $p_0=13$, the evolution of phase differences over time for different scale parameters $\xi$ is shown. The left panel displays the radial phase difference, and the right panel shows the azimuthal phase difference, where the phase difference is defined as $\Delta\Phi_i=\left|\Phi_i(\xi\neq0)-\Phi_i(\xi=0)\right|$. The red horizontal line in the figures represents the phase threshold that can be distinguished by detection.}
\label{d}
\end{figure*}

\begin{figure*}[]
\includegraphics[width=1 \textwidth]{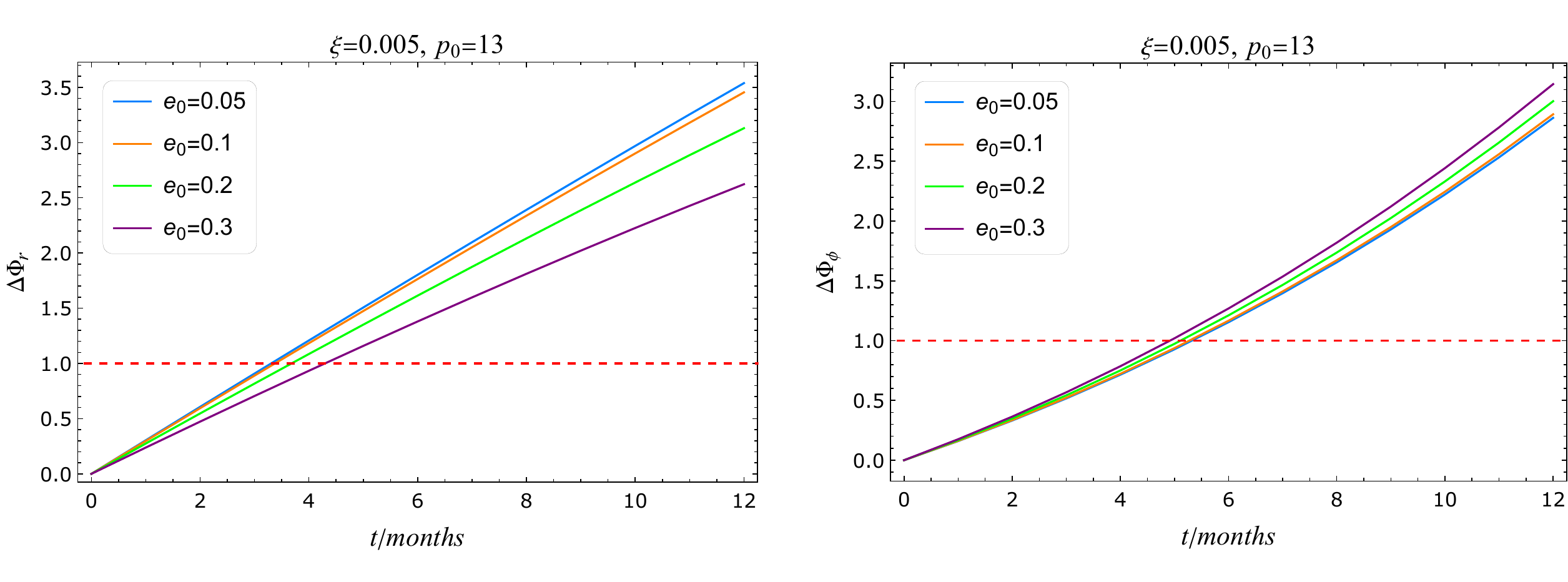}
\caption{
Under fixed scale parameter $\xi=0.005$, the evolution of phase differences over time for different eccentricities $e_0$ is shown. The left panel displays the radial phase difference, and the right panel shows the azimuthal phase difference, where the phase difference is defined as $\Delta\Phi_i=\left|\Phi_i(\xi\neq0)-\Phi_i(\xi=0)\right|$. The red horizontal line in the figures represents the phase threshold that can be distinguished by detection.}
\label{e}
\end{figure*}

In Figs. \ref{d} and \ref{e}, we calculate the phase difference between the regular black hole and the Schwarzschild black hole by numerically solving equation (\ref{31}), defined as $\Delta\Phi_i=\left|\Phi_i(\xi\neq0)-\Phi_i(\xi=0)\right|$, with all initial phases set to $\Phi_i|_{t=0}=0$. 
It is worth noting that we adopt the method of fixing initial orbital parameters $(p_0, e_0)$ for phase evolution. This differs from the frequency matching technique employed by Warburton et al. \cite{Warburton:2011fk} when comparing different approximation methods within the same Schwarzschild spacetime. Warburton et al.'s work represents an important contribution to the field, where their frequency matching method effectively eliminates artificial phase offsets caused by computational approximation differences, thereby accurately extracting physical effects. In our study, since we are comparing two spacetimes with different geometric structures, the orbital frequency differences induced by the asymptotic safety gravity parameter $\xi$ precisely carry the quantum correction information we seek to detect. Similar fixed orbital parameter approaches for phase evolution have been widely adopted in EMRIs studies within modified gravity theories and loop quantum gravity, as seen in references \cite{Fu:2024cfk, Zi:2024jla, Zi:2023qfk, Zhang:2022rfr, Kumar:2024our, Kumar:2024utz, Zi:2025idv, Datta:2019epe}, among others.

Fig. \ref{d} fixes the initial conditions $e_0=0.1$ and $p_0=13$, showing the evolution of phase differences over time under different scale parameters; Fig. \ref{e} fixes the scale parameter $\xi=0.005$ and $p_0=13$, showing the time evolution of phase differences under different initial eccentricities. According to the literature \cite{Datta:2019epe,Zi:2023omh,Zi:2023qfk}, when the phase difference exceeds $1$ radian (black horizontal line in the figures), LISA can effectively distinguish between these two signals. Evidently, both the radial and azimuthal phase differences increase monotonically with time, eventually exceeding LISA's detection threshold. Fig. \ref{d} indicates that under fixed initial eccentricity conditions, larger scale parameters lead to more rapid growth in phase differences. Even when the scale parameter is only of the order of $10^{-3}$, the phase difference after several months can reach LISA's resolution threshold, providing a preliminary basis for constraining the scale parameter. Fig. \ref{e} shows that under fixed scale parameter conditions, the initial eccentricity has a relatively minor impact on the phase difference. Overall, the influence of the scale parameter on the phase difference is significantly greater than that of the initial eccentricity.

\begin{figure*}[]
\includegraphics[width=1 \textwidth]{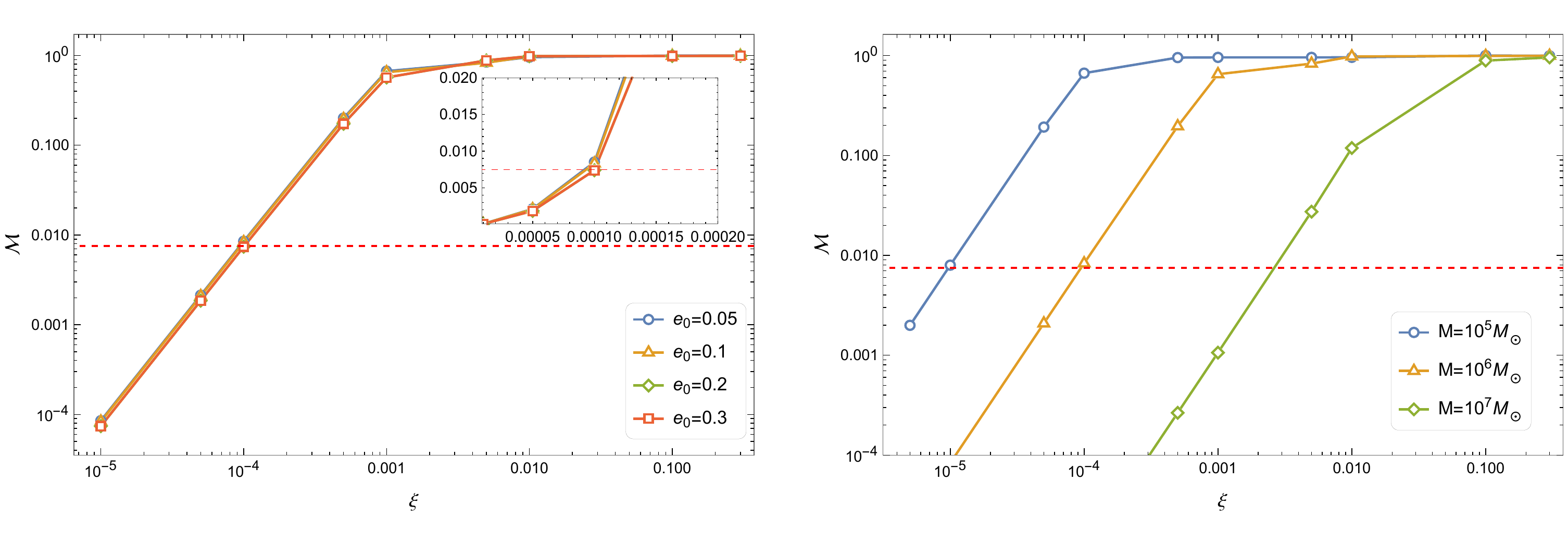}
\caption{
With one year of evolution time, the figure shows the waveform mismatch between the regular black hole in asymptotically safe gravity and the Schwarzschild black hole. The left panel displays the waveform mismatch under fixed other parameters for eccentricities $e_0\in\{0.05,0.1,0.2,0.3\}$; the right panel shows the waveform mismatch under fixed other parameters for masses $M\in\{10^5M_\odot,10^6M_\odot,10^7M_\odot\}$. The red dashed line represents LISA detector's minimum identification threshold $\mathcal{M}=0.0075$.}
\label{f}
\end{figure*}

Fig. \ref{f} quantitatively demonstrates the mismatch between the waveforms of the regular black hole in asymptotically safe gravity and the traditional black hole, with an observation time of one year. The results indicate that as the scale parameter $\xi$ increases, the waveform mismatch gradually increases and eventually exceeds the identification threshold of the LISA detector (the red dashed line represents LISA's detection threshold). At larger scale parameters, the mismatch approaches $1$. The left panel of Fig. \ref{f} shows that under fixed other parameters, the effect of eccentricity on the mismatch is relatively small, which is consistent with the phase difference analysis results in Fig. \ref{e}. The right panel of Fig. \ref{f} displays the mismatch under different masses of the primary black hole, indicating that systems with smaller mass black holes more easily reach the detection threshold. Comprehensive analysis shows that for systems with a mass of $10^6M_\odot$, LISA's detection capability for the scale parameter $\xi$ during a one-year observation period can reach the order of $\sim 10^{-4}$, suggesting that the constraint precision may also approach this magnitude, a precision that cannot be achieved by other observational methods. 
For a system with $M = 10^6 M_{\odot}$, the characteristic scale $L_{\xi} \approx 1.477 \times 10^7 \text{ m}$ corresponding to this detection threshold is much larger than the Planck scale $L_p \approx 1.616 \times 10^{-35} \text{ m}$ ($L_{\xi}/L_p \approx 9\times 10^{41}$). This scale difference indicates that the quantum gravitational effects we can detect are far above the Planck scale. Nevertheless, compared to macroscopic scales such as the Schwarzschild radius, $L_{\xi}/r_s \approx 5 \times 10^{-3}$ still represents a small correction. During the long-term evolution of EMRIs, small-mass objects complete thousands to tens of thousands of orbital cycles in the strong-field region, enabling these tiny quantum gravitational effects to reach detectable levels through continuous phase accumulation, thereby leaving observable characteristic signatures in gravitational wave signals. This mechanism provides a possible pathway for constraining quantum gravity parameters through space-based gravitational wave observations.

Furthermore, to further validate the reasonableness of the scale parameter $\xi$ values appearing at the 3PN order mentioned above, we compare with the observational results from LIGO-Virgo-KAGRA GWTC-3 \cite{LIGOScientific:2021sio}. We adopt an analysis method similar to that in the literature \cite{Kumar:2025njz,AbhishekChowdhuri:2023gvu}. By combining equations (\ref{29}) and (\ref{31}), we obtain the phase deviation $\Delta\Phi = f(e,q,M,\Omega, \xi)$ in the frequency domain. Given the complexity of the complete analytical expression, we omit its explicit form here, but the expression can be obtained directly through computer symbolic solving. For simplicity, we set the eccentricity $e_{in} = 0$, and according to the literature \cite{LIGOScientific:2021sio}, we select a binary black hole system with $M \sim 140M_{\odot}$ and mass ratio $q \sim 0.6$ for order-of-magnitude estimation. For the scale parameter range $\xi \in (10^{-4}, 10^{-1})$, combined with the frequency range shown in Table V of the observational literature \cite{LIGOScientific:2021sio}, our order-of-magnitude analysis indicates that the phase deviation $\Delta\Phi$ lies within the range of $10^{-8}-10^{-2}$. This computational result is consistent with the 3PN order deviation $\delta\phi_6 \sim 0.16$ given by GWTC-3 observational data (see Table VI in \cite{LIGOScientific:2021sio}), indicating that the scale parameter values considered in this work possess a certain degree of reasonableness.

Meanwhile, the results of this study exhibit remarkable quantitative consis tency with independent studies based on different quantum gravity theoretical frameworks. For example, research within the loop quantum gravity framework \cite{Fu:2024cfk, Zi:2024jla} predicts that LISA detectors can identify quantum gravity effects on the order of $10^{-4} \sim 10^{-6}$, while the LISA detection threshold $\xi \sim 10^{-4}$ obtained in this work based on asymptotic safety gravity theory is highly consistent with these results in terms of order of magnitude. This quantitative consistency across theoretical frameworks provides cross-validation for our conclusions and enhances their credibility to a considerable extent.

\subsection{Constraints on the Scale Parameter}

To further quantify the measurement precision of the scale parameter, in this subsection we employ the FIM method for parameter uncertainty analysis. The FIM provides the optimal statistical precision attainable under given observational conditions and is an effective tool for evaluating parameter constraint capabilities. To ensure the robustness of the analysis results, we control the signal-to-noise ratio at approximately $50$ by adjusting the system's luminosity distance, which both guarantees sufficient statistical significance of the signal and avoids non-Gaussian effects that might be introduced by excessively high signal-to-noise ratios. In our calculations, we set the scale parameter to $\xi=0$ (i.e., the value expected in general relativity), while fixing other physical parameters (such as $M$, $m$, $p_0$, and $e_0$).

\begin{figure}[]
\includegraphics[width=1 \textwidth]{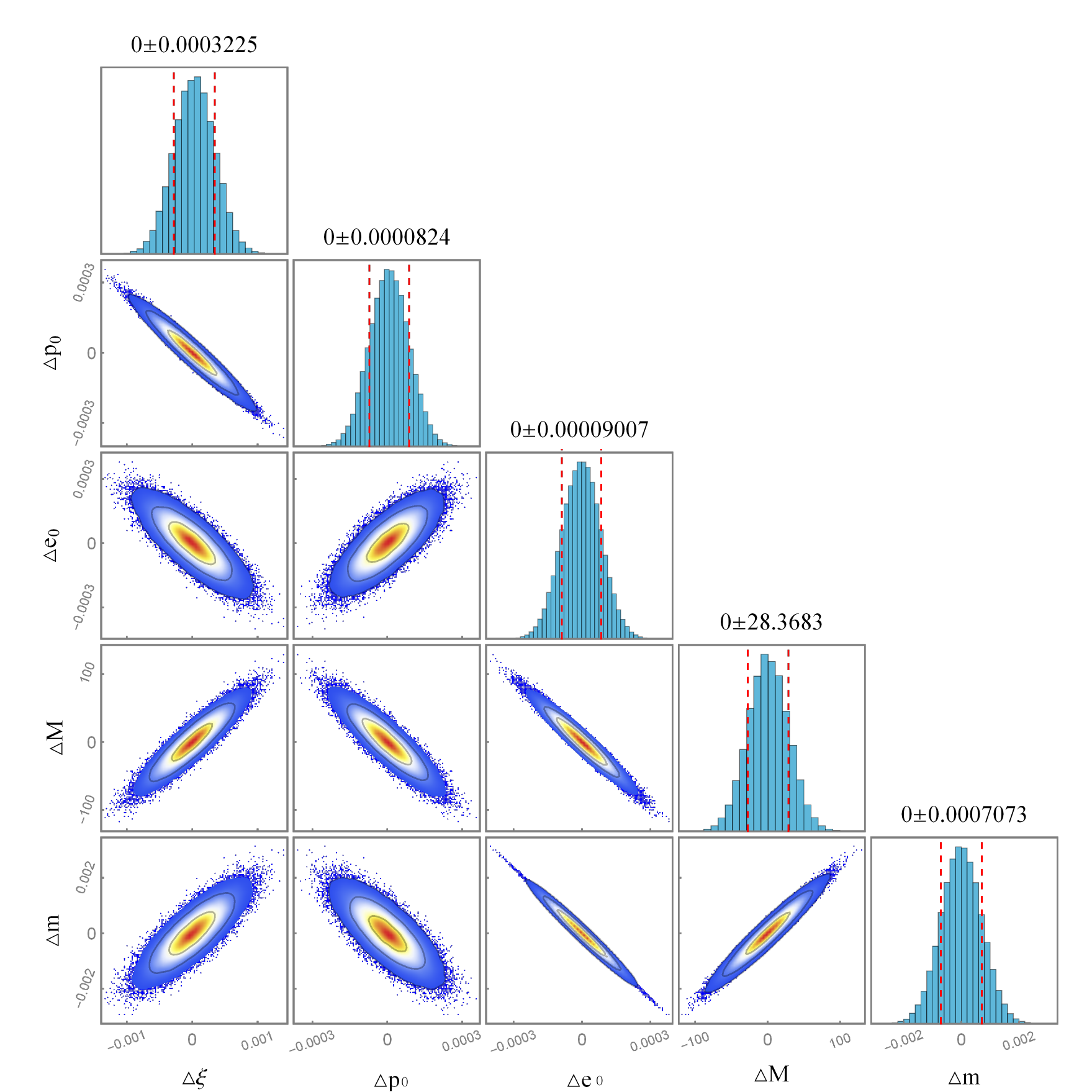}
\caption{
Probability distribution diagram of intrinsic parameters in extreme mass ratio inspiral systems. The scale parameter $\xi=0$, red dashed lines indicate the first confidence interval, and contours from outer to inner correspond to $99\%$, $95\%$, and $68\%$ probability distribution regions, respectively.
}
\label{g}
\end{figure}

Fig. \ref{g} displays the joint posterior probability distribution across multiple parameters, including the correlation structure between the scale parameter $\xi$ and other key parameters, as well as their marginal probability distributions. The analysis indicates that within a one-year LISA observation period, 
the theoretical $1\sigma$ measurement precision for the scale parameter can reach $\Delta\xi\approx3.225\times10^{-4}$.
This constraint precision is at a level that cannot be achieved by other existing observational methods. Therefore, gravitational wave signals produced by extreme mass ratio inspirals possess unique advantages in testing the characteristics of regular black holes within the asymptotically safe gravity framework.

To evaluate the numerical reliability of the above results, we conducted systematic stability tests following the standard methods in the literature \cite{Piovano:2021iwv}. Specifically, we applied perturbations of two scales, namely $\pm10^{-6}$ and $\pm10^{-9}$, respectively, to evaluate the stability of the system.  The results show that under $\pm10^{-6}$ scale perturbations, the system stability is $4.364\times10^{-4}$; under $\pm10^{-9}$ scale perturbations, the system stability is $4.386\times10^{-7}$. These systematic test results strongly support the reliability and robustness of our main analytical conclusions.

\section{\label{sec5}Conclusion}

This paper investigates the regular black hole model proposed by Bonanno et al. \cite{Bonanno:2023rzk} within the asymptotically safe gravity framework. This model modifies the Misner-Sharp mass function to $M\left(r\right)=\frac{r^3}{6\xi}\ln\left(1+\frac{6M_0\xi}{r^3}\right)$ by introducing a gravity-matter coupling function $\chi $, thereby eliminating the singularity at the center of classical black holes. The scale parameter $\xi$ cannot be precisely determined from first principles, thus requiring constraints from astronomical observations. Based on this, we used the AAK method to approximately calculate the gravitational wave characteristics produced by this regular black hole model in EMRIs  on the equatorial plane. By quantifying the phase accumulation shift caused by quantum gravity effects, we assessed the potential constraint capability of gravitational wave observations on $\xi$, providing theoretical basis for verifying the effectiveness of asymptotically safe quantum gravity theory in the strong-field regime.

Specifically, we rigorously derived the energy and angular momentum fluxes, fundamental orbital frequencies, and orbital evolution equations on the equatorial plane. To evaluate the influence of the scale parameter, we rigorously calculated the leading-order contribution of $\xi$, finding that the effects of $\xi$ are primarily manifested in higher-order terms. Subsequently, we constructed complete gravitational wave waveform templates using the AAK method.

Through systematic comparison of phase differences between waveforms of this model and Schwarzschild black hole, we quantitatively assessed the constraint capability on the scale parameter. We constructed gravitational wave waveforms for scale parameters $\xi=0$ (Schwarzschild  black hole), $\xi=0.01$, and $\xi=0.001$ under different initial eccentricity conditions. Results show that after several months of observation period, the waveforms exhibit significant differences, with larger $\xi$ values highlighting differences earlier. This conclusion was rigorously verified in phase deviation analysis: larger $\xi$ values more easily exceed the phase detection threshold, while eccentricity has no significant impact on phase deviation.

Additionally, by calculating the mismatch between waveforms with scale parameter corrections and Schwarzschild black hole waveforms, good constraints on $\xi$ can be obtained within the LISA detection threshold. For example, for EMRIs with $M=10^6M_\odot$, LISA gravitational wave observations can detect scale parameters $\xi$ on the order of $\sim10^{-4}$, a constraint precision significantly superior to all other current astronomical observation methods. Finally, we applied the FIM method to rigorously evaluate the measurement error of $\xi$. Results indicate that under $M=10^6M_\odot$, the measurement precision of LISA detector for $\xi$ is $\Delta\xi\approx3.225\times10^{-4}$, providing observable experimental verification evidence for asymptotically safe gravity theory.

This study preliminarily explores the potential of using EMRIs to constrain the scale parameter $\xi$ in asymptotic safety gravity theory. We employ the Peters and Mathews quadrupole approximation method \cite{Peters:1963ux,Peters:1964zz} to calculate gravitational wave radiation flux, and find that asymptotic safety gravity corrections introduce contributions from the scale parameter $\xi$ at the 3PN order in the post-Newtonian expansion. It should be noted that the analytical framework in this work is based on the quadrupole approximation and does not incorporate the complete content of full 3PN order general relativistic post-Newtonian corrections. Specifically, this approximation method primarily retains quadrupole radiation contributions, while the complete 3PN theory also includes higher-order multipole radiation effects, more precise orbital dynamics corrections, and other contributions. These complete 3PN order general relativistic contributions may alter the quantitative assessment of the detectability of parameter $\xi$. Therefore, the detection threshold $\xi \sim 10^{-4}$ obtained in this work under the current approximation framework should be understood as an order-of-magnitude estimate. A self-consistent analysis method incorporating complete 3PN order general relativistic corrections will be developed in detail in future work, which will provide more reliable theoretical predictions for precise constraints on the scale parameter of asymptotic safety gravity theory.

\section{acknowledgements}
We acknowledge the anonymous referee for a constructive report that has significantly improved this paper. This work was  supported by Guizhou Provincial Basic Research Program(Natural Science)(Grant No. QianKeHeJiChu-[2024]Young166),  the Special Natural Science Fund of Guizhou University (Grant No.X2022133), the National Natural Science Foundation of China (Grant No. 12365008) and the Guizhou Provincial Basic Research Program (Natural Science) (Grant No. QianKeHeJiChu-ZK[2024]YiBan027) .

\appendix
\section{Energy Flux and Angular Momentum Flux Derivation}\label{appA}
In this appendix, we start from the classical flux formula derived by Peters and Mathews \cite{Peters:1963ux,Peters:1964zz} and provide a detailed derivation of the gravitational wave flux expression required in this paper.

The flux formula derived by Peters and Mathews  is
\begin{equation}
\left\langle\frac{dE}{dt}\right\rangle=\frac{1}{5}\mu\left\langle\frac{d^3Q_{ij}}{dt^3}\frac{d^3Q^{ij}}{dt^3}-\frac{1}{3}\frac{d^3Q_{ii}}{dt^3}\frac{d^3Q^{jj}}{dt^3}\right\rangle,
\label{A1}
\end{equation}

\begin{equation}
\left\langle\frac{dL_i}{dt}\right\rangle=\frac{2}{5\mu M}\epsilon_{ijk}\left\langle \frac{d^2Q_{jm}}{dt^2}\frac{d^3Q^{km}}{dt^3}\right\rangle.
\label{A2}
\end{equation}
The moment of inertia tensor $Q_{ij}$ is
\begin{equation}
Q_{ij}=\mu x^ix^j.
\label{A2}
\end{equation}
Here, $x^i$ represents the position vector between the stellar-mass CO and the regular black hole in asymptotically safe gravity.

To study the effect of the scale parameter on gravitational wave radiation, we only calculate up to the leading order term in the scale parameter. Therefore,
the average energy flux can be expressed as
\begin{align}
\left\langle \frac{dE}{dt}\right\rangle  
&= \frac{X^3(37e^{4}+292e^{2}+96)\mu^{2}}{15M^{2}p^{5}}+\frac{e^{2}X^3(53e^{4}+450e^{2}+176)\mu^{2}}{5M^{2}p^{6}}\nonumber\\
&+\frac{X^3\left[2160e^{8}+(456 - 5880X)e^{2}-2880(X-1)+(1110X+13559)e^{6}+18(425X-728)e^{4}\right]\mu^{2}}{60M^{2}p^{7}}\nonumber\\
&+\frac{F(\xi,e)\mu^2}{M^2p^{8}}+O(1/p^{9}).
\label{A4}
\end{align}
The average angular momentum flux is
\begin{align}
\left\langle\frac{dL}{dt}\right\rangle   
&= \frac{4X^3(8 + 7e^{2})\mu^{2}}{5Mp^{7/2}}+\frac{4e^{2}X^3(38 + 27e^{2})\mu^{2}}{5Mp^{9/2}}\nonumber\\
&+\frac{3X^3\left[207e^{6}+4(5X - 33)e^{2}-160(X - 1)+(140X + 43)e^{4}\right]\mu^{2}}{10Mp^{11/2}}+\frac{G(\xi,e)\mu^2}{M p^{13/2}}+O(1/p^{15/2}).
\label{A5}
\end{align}
Here, $X=\sqrt{1-e^2}$, where $F(\xi,e)$ and $G(\xi,e)$ are correction functions containing the scale parameter, with specific expressions given by
\begin{align}
F(\xi,e)&=X^3\left\{108 e^{10}+\frac{502739 e^8}{1280}-\frac{17104 e^7}{275}-\frac{71469783 e^6}{12800}+\frac{1757936e^5}{1225}-\frac{177180337 e^4}{6144}\right.\nonumber\\
&+\frac{764430272 e^3}{30625}-\frac{971493 e^2}{32}+\frac{14148672e}{625}-\frac{23037}{5}+X \left(135 e^8+1015 e^6-450 e^4-604 e^2-96\right)\nonumber\\
&+\pi ^2 \left(\frac{3093 e^6}{20}+\frac{384633 e^4}{80}-\frac{27072 e^3}{25}+14019 e^2-\frac{19008
   e}{25}+\frac{18144}{5}\right)\nonumber\\
   &+\left.\left[\frac{337 e^6}{10}+\frac{4084 e^4}{5}+\frac{10748 e^2}{5}+X\left(-\frac{37 e^6}{5}-51
   e^4+\frac{196 e^2}{5}+\frac{96}{5}\right)+\frac{2208}{5}\right] \xi\right\},
\label{A6}
\end{align}
\begin{align}
G(\xi,e)&=X^3\left\{\frac{1863 e^8}{10}-\frac{12 e^7}{5}-\frac{99351 e^6}{640}-\frac{57324 e^5}{175}-\frac{5360461
   e^4}{1280}-\frac{1763028 e^3}{6125}-\frac{51159 e^2}{4}\right.\nonumber\\
&+\frac{1244388 e}{125}-\frac{18069}{5}+X \left(288 e^6+168 e^4-360 e^2-96\right)+\pi ^2 \left(\frac{3231 e^4}{10}+\frac{24528 e^2}{5}+\frac{17568}{5}\right)\nonumber\\
&\left.+\left[636 e^2+X \left(-\frac{84 e^4}{5}-\frac{12
   e^2}{5}+\frac{96}{5}\right)+\frac{1824}{5}\right] \xi\right\}.
\label{A7}
\end{align}

Furthermore, to clarify the specific order of the scaling parameter $\xi$ in the Post-Newtonian (PN) expansion, we rewrite expressions \ref{A4} and \ref{A5} as:
\begin{align}
\frac{dE}{dt}= \frac{\mu^2}{ M^2 p^5}\left[a_1(e) + \frac{a_2(e)}{p} + \frac{a_2(e)}{p^2} + \frac{a_3(e)}{p^3} + \frac{a_4(e)\xi}{p^3}+ \cdots\right] ,
\label{A8}
\end{align}
\begin{align}
\frac{dL}{dt} = \frac{\mu^2 }{Mp^{7/2}}\left[b_1(e) + \frac{b_2(e)}{p} + \frac{b_2(e)}{p^2} + \frac{b_3(e)}{p^3}+ \frac{b_4(e)\xi}{p^3} + \cdots\right].
\label{A9}
\end{align}
where the corresponding coefficients $a_1(e)$, $a_2(e)$, $a_3(e)$, $a_4(e)$, $b_1(e)$, $b_2(e)$, $b_3(e)$, $b_4(e)$ can be read from expressions \ref{A4}, \ref{A5}, \ref{A6}, \ref{A7} respectively.

Through the above expansion, it can be clearly seen that the scaling parameter $\xi$ only begins to contribute at the 3PN order. That is, the scaling parameter $\xi$ is introduced as a correction term to the standard 3PN order results of general relativity, indicating that our analysis preserves the lower-order PN terms of general relativity unchanged while introducing new physical effects only at the 3PN order.

\bibliography{ref}

\end{document}